\def \yskip{\penalty-50\vskip3pt plus 3pt minus 2pt}
\def \reference {\par \yskip \noindent \hangindent .4in \hangafter 1}
\def \abc#1#2#3#4 {\reference{} #1, {\sl#2}, {\bf#3}, #4}
\def \blank {\lower 5pt\hbox to 0.75in{\hrulefill}}

\def \K{\rm{K}}

\def \yr{\rm{yr}}

\def \s{~\rm{s}}

\def \kpc{~\rm{kpc}}
\def \km{~\rm{km}}
\def \lae{\mathrel{<\kern-1.0em\lower0.9ex\hbox{$\sim$}}}
\def \gae{\mathrel{>\kern-1.0em\lower0.9ex\hbox{$\sim$}}}
\def \yskip{\penalty-50\vskip3pt plus 3pt minus 2pt}
\def \reference {\par \yskip \noindent \hangindent .4in \hangafter 1}
\def \abc#1#2#3#4 {\reference{} #1, {\sl#2}, {\bf#3}, #4}
\def \blank {\lower 5pt\hbox to 0.75in{\hrulefill}}

\def \yr{~\rm{yr}}
\def \yrs{~\rm{yrs}}

\def \erg{~\rm{erg}}
\def \s{~\rm{s}}
\def \km{~\rm{km}}

\def \lae{\mathrel{<\kern-1.0em\lower0.9ex\hbox{$\sim$}}}
\def \gae{\mathrel{>\kern-1.0em\lower0.9ex\hbox{$\sim$}}}

\documentstyle[11pt,aasms,tighten,flushrt]{article}

\begin{document}
\small

\setcounter{page}{1}

\begin{center}
\bf 
A MODERATE CLUSTER COOLING FLOW MODEL 
\end{center}

\begin{center}
Noam Soker\\
Department of Physics, University of Haifa at Oranim\\
Tivon 36006, Israel; soker@physics.technion.ac.il \\
\medskip
Raymond E. White III\\
Code 662, NASA Goddard Space Flight Center, Greenbelt, MD 20771\\
and Department of Physics \& Astronomy, University of Alabama,
Tuscaloosa, AL 35487-0324; white@merkin.astr.ua.edu\\
\medskip
Laurence P. David, and Brian R. McNamara $^{(1)}$ \\
Harvard Smithsonian Center for Astrophysics \\
60 Garden Street, Cambridge, MA 02138 \\
lpd@head-cfa.harvard.edu; brm@head-cfa.harvard.edu \\
$^{1}$ also:
Dept. of Physics \& Astronomy, Ohio University \\
Clippinger Lab 339, Athens, OH 45701 
\end{center}
%
\begin{center}
\bf ABSTRACT
\end{center}

 We propose that the outer portions of cooling flows
in clusters of galaxies are frequently disrupted by radio jets,
and that their {\it effective} ages are much shorter than
the cluster ages.
 The inner regions, where the gas density is higher, are more difficult to
disrupt, and may continue to harbor cooling flows even after 
disruption events. 
 The main assumption of the proposed scenario is that, on a time scale of
$\sim 2-4 \times 10^9 \yrs$, the cD galaxies in cooling flow clusters
undergo powerful bursts of AGN activity that produce strong radio jets.
 The radio jets excite shocks in the inner regions
($r \lesssim 100 \kpc$) of cooling flow clusters.
 A radio burst may result from the accretion of cooling material 
by the central black hole or from a collision with a subcluster.
 We assume that the jets remain strong, with kinetic powers of
$\sim 10^{47} \erg \s^{-1}$ for $\sim 10^7$ years.
 The jets excite shock waves moving at several$\times 10^3 \km \s^{-1}$
and heat the cooling flow region, hence terminating it in the outer regions.
 The proposed scenario predicts that the total accreted mass due to the
 cooling flow is an order of magnitude lower than the mass accreted
according to the ``standard'' cooling flow model (which assumes
an undisturbed cooling flow for a time equal to the age of the cluster).
 The scenario, therefore, brings to agreement the observations that a
large fraction of clusters harbor cooling flows,
that strong optical and radio activity are present only in the very
inner regions of cooling flows, and the lack of a satisfactory reservoir
of the expected (in the standard cooling flow model)
large mass that has been cooling over the life of the cluster.

\noindent {\it Subject headings:} 
galaxies: clusters: general. 
galaxies: cooling flows

\clearpage 

\section{INTRODUCTION}

 At the centers of most galaxy clusters, the radiative cooling time
of intracluster gas is substantially shorter than the age of the Universe.
As the central gas cools and loses pressure support, the weight of 
the surrounding gas compresses it, causing it to flow inward in a 
``cooling flow'' (Fabian 1994 and references therein). 
There are two major problems yet to be solved before the cooling flow model
is widely accepted. 
First, there is no direct observational indication for inflow
in cooling flows. 
 Second, and more severe, the final reservoir of the cooling gas 
has not been detected (for a review see O'Dea \& Baum 1997).
 
Inflow velocities are expected to be subsonic, small enough that the
spectral resolution required to detect such velocities is beyond the
capabilities of the current generation of X-ray satellites.
The second problem, determining the ultimate sink for cooling gas,
has always been the more fundamental difficulty 
in the cooling flow scenario.
However, no other (stable) model has been proposed to account for the short 
cooling times of the hot intracluster medium as inferred from X-ray 
observations. 
Moreover, many cluster properties are unique to cooling flow clusters:
(1) Prominent optical filaments are found only in clusters having large 
X-ray inferred cooling rates.
(2) Strong intracluster magnetic fields, as inferred from large
Faraday rotation measures, are found only in cooling flow clusters
(e.g., Dreher, Carilli, \& Perley 1987; Taylor {\it et al.} 1990; 
Ge \& Owen 1993, 1994; Carvalho 1994), and can 
be explained by the gaseous compression of cooling flows  (Soker \& Sarazin 
1990; 
Christodoulou \& Sarazin 1996; Soker 1997; Godon, Soker \& White 1998;
Goncalves \& Friaca 1999).
(3) Radio jets in many cooling flow clusters are disrupted a few kpc from 
the centers of the cD galaxies from which they emerge;
 although the disruption of the jets via collision with cold gas has
been demonstrated in specific cases (McNamara {\it et al.} 1996), 
the disruptions may be explained by the presence of cooling flows
as well (Soker \& Sarazin 1988; Sumi, Norman, \& Smarr 1988;
Sulkanen, Burns, \& Norman 1989; Loken {\it et al.} 1993). 
(4) The central regions of cD galaxies
embedded in cooling flows have anomalously blue colors
which are correlated with the cooling rates
inferred from X-ray observations (see next section).
(5) Cooling flow clusters may have a central excess of heavy elements.
One common feature in all of these observations is that all 
activity observed at wavelengths longer than X-rays is confined
to the central $\lesssim 30 \kpc$ of cooling flow clusters,
which is much smaller than the typical cooling radius, 
(i.e., the radius where the cooling time equals the cluster age).
It therefore seems as if long lived cooling flows only exist in the 
very inner regions of clusters. 

In the present work, we propose that the outer portions of cooling flows 
are frequently disrupted by radio jets, and that the  
{\it effective} ages of cooling flows are much smaller than the cluster ages. 
 That cooling flows may be disrupted by activity in the nuclei
of the cD galaxies, rather than by subcluster mergers, was suggested by
Allen {\it et al.} (2000; hereafter AFJAN), although they did not
consider radio jet outbursts.
 The outburst of transient radio jets, blown by central black holes, may be
triggered by subcluster mergers (Bliton {\it et al.} 1998),
or by accretion of the cooling material, producing a duty cycle. 
The inner regions, where the gas density is highest, are more difficult to
disrupt, and may continue to harbor cooling flows even after a
radio outburst.
However, the outer regions ($\gtrsim 30 \kpc$) must start a new cooling flow
after the disruption.
This scenario accounts for:
(i) The small total mass accreted into the central galaxy by the cluster cooling 
flow.
(ii) The presence of multiwavelength activity within the central region, as
discussed above. 
(iii) Observations that many clusters possess cooling flows 
(since the cooling flow in the inner region is not disturbed,
or it restarts again very quickly). 
{{{
In previous analogous work on $galactic$ (rather than cluster)
cooling flows, Binney \& Tabor (1995) suggested that cooling 
flows in elliptical galaxies are cyclical: cooling flow accretion
onto a massive central black hole generates
nonrelativistic jets which suppress further accretion;
after the accretion is suppressed, the jets are no longer
powered, so they fizzle, allowing the accretion to resume. 
Ciotti \& Ostriker (1997) subsequently described a similar model,
where galactic cooling flows are cyclically suppressed by 
radiation from the accretion disks around massive central black holes,
rather than by jet interactions.
}}}

Our ``moderate'' cooling flow model, while 
reducing the impact of cooling flows, puts it on a more solid ground 
observationally.
In $\S 2$ we summarize the relevant observations to support the proposed
scenario, while in $\S 3$ we present the basic scenario.
In $\S 4$ we discuss the evolution with time of an undisturbed, as
well as disturbed, cooling flow, and show that the total accumulated mass
is an order of magnitude less than that in models which assume
cooling flows ages equal to clusters' ages.
 In $\S 5$ we discuss our results and compare them to the recent
paper by Allen {\it et al.} (2000).
We summarize our main results in $\S 6$.

\section {OBSERVATIONAL CONSIDERATIONS}

Essentially all signatures of accretion (i.e.  gas with $T\lae 10^7$K
and star formation) are found in the inner few
to $\sim$10\% of the cooling region of clusters.  The most prominent
signature is the strong, spatially extended nebular line emission 
observed in roughly half of cluster cooling flows 
(Heckman 1981; Cowie {\it et al.} 1983;
Hu, Cowie \& Wang 1985; Heckman {\it et al} 1989; see
Baum 1992 for a review).  The origin and nature of the heating
and ionization mechanism(s) for the $T=10^4$ K gas are 
critical to our understanding of cooling flows, and these
questions have been pursued vigorously since the discovery 
of cooling flows (Fabian \& Nulsen 1977; Cowie \& Binney 1977).
Several heating mechanisms are probably important, including shocks,
photoionization from young stars, and photoionization by soft X-ray
emission from the surrounding hot gas (Voit, Donahue \& Slavin 1994).
The optical line luminosities are too large by one to two orders of magnitude
to be powered by simple recombination of cooling gas.
However, the unique association of line emission with cooling flows
suggests that the $T\sim 10^4$ K gas has condensed out of the 
cooling flow or has been deposited by processes related to
the presence of a cooling flow.

Searches for the 21 cm line of neutral hydrogen (H I) toward 
cooling flow central dominant galaxies
 (McNamara, Bregman, \& O'Connell 1990; O'Dea, Gallimore, \& Baum
1995) have resulted in only a few detections of H I 
in absorption against the central radio sources.
In two cases the H I is spatially extended
over at most a few kpc (O'Dea, Baum, \& Gallimore 1994; Taylor
1996).   Although it is certainly possible
that clouds of low column density H I are present at larger radii, the
columns must be much less than the columns found in
X-ray measurements of soft X-ray absorption, or the gas would have
been detected against the outer lobes of large radio sources 
such as Hydra A (Taylor 1996). 

The $2.1\mu$ infrared molecular hydrogen (H$_2$) feature has been 
detected at the centers of essentially all cooling flows with
strong nebular lines (e.g. H$\alpha$) emission (Elston \& Maloney
1992; Jaffe, \& Bremer 1997; Donahue et al.~2000). 
This gas is relatively cold, with a temperature $T\sim 2000$ K, and 
is concentrated within the inner several kpc of the central cluster
galaxies.  Its properties are consistent with being the dense gas
in the inner regions of the clouds responsible for 
the optical line emission.
 The total mass of the $2000$ K clouds is poorly known.
 Searches for the carbon monoxide (CO) molecule in the millimeter
$1\rightarrow 0$ and $2\rightarrow 1$ transitions has yielded emission
upper limits of $\sim 10^{8-9}M_\odot$ 
(McNamara, \& Jaffe 1994; O'Dea {\it et al.} 1994; Braine \& Dupraz 1994).
CO emission has been detected in NGC 1275,
the central galaxy of the Perseus cluster (Lazareff et al. 1989);
however, this is the only CO detection in a cooling flow,
and the relationship of the $\sim 10^{10} M_\odot$ of molecular
gas to the cooling flow is not understood.

Central dominant galaxies in cooling flows are well-known to be
unusually blue in their central regions (see McNamara 1997, 1999 for reviews).  
The incidence of unusually blue colors in a central dominant galaxy,
as well as the magnitude of the color excess, are correlated with the
total cooling rate of the intracluster gas (McNamara 1997; Cardiel, Gorgas, 
\& Aragon-Salamanca 1998).  Although
there are several possible origins for the blue light
(e.g., star formation, scattered light from an active nucleus,
nebular continuum, a metal poor stellar population), 
the preponderance of evidence points to populations of
young stars in most instances.  The correlation between 
blue color excess and cooling rate is consistent with the
blue light being associated with a young population
of stars that has been fueled by accreted gas from the cooling
flow (Allen 1995; McNamara 1997, 1999; Cardiel {\it et al.} 1998).
However, the accretion populations are confined to the
inner 5--30 kpc of the central cluster galaxies. 
The young populations, as traced by the blue color excesses, 
are extended, in general, over similar 
spatial scales as the emission-line regions and the radio sources 
(Cardiel {\it et al.} 1998).
On the other hand, the cooling regions in the standard model
(where the age of the cooling flow is assumed to be equal
to the age of the cluster) extend to $\sim 100-150$ kpc,
a region that is more than an order of magnitude larger in radius
compared to the blue regions.  

In order to reproduce the observed X-ray surface brightness
profile in cooling flows, the mass accretion rate must 
vary as $\dot M \propto r$ (e.g., Thomas, Fabian \& Nulsen 1987).
Such a radial dependence indicates that the cooling gas
is deposited over a broad region, with most of the gas deposited near the
cooling radius.
In fact, the star formation rates inferred from the blue color excess
are consistent (within a factor or 2 or 3) with the mass deposition
rates within the central 10~kpc
(Allen 1995; McNamara 1997, 1999; Cardiel {\it et al.} 1998). 
However, the fate of the remaining $90\%$ of the gas
that is deposited beyond the central 10~kpc is still unknown.

\section {THE PROPOSED SCENARIO}

The primary mechanism of our ``moderate" cooling flow scenario is 
that cooling flow clusters undergo intermittent very powerful radio
outbursts, i.e., their central black holes blow very strong
radio jets.
 An outburst may be triggered by the accretion of cooling flow material
or by a subcluster merger. 
 The intermittent mergers scenario may get support from a
recent paper by Loken, Melott, \& Miller (1999), who find that
cooling flow clusters with high mass accretion rates reside in more
crowded environments.
 An example of a cooling flow cluster which is going through a merging
event is the Centaurus cluster (Churazov {\it et al.} 1999).
These mergers can affect the cooling flow in several ways:
 (1) Shock waves may be excited directly, or
 (2) may result from induced AGN activity, 
      including strong radio jets, or from induced starbursts which
      will lead to a burst of supernovae.
      The radio jets will also excite shock waves in the intracluster gas, 
      which may
      turn off the cooling flow in the outer regions, but probably
      not in the inner regions as we discuss below. 
 (3) Subcluster mergers increase the turbulence in the hot gas, and may
     increase the angular momentum of the intracluster gas near the
     center of the cD galaxy. This in turn may enhance magnetic activity
     (Godon {\it et al.} 1998).
 (4) The merger may increase the central mass concentration as
     low entropy gas settles into the center of the merged remnant. 
  This will produce a denser intracluster gas in the inner region, decreasing the cooling
  time.  The main effect in this case is that a cluster that does not yet have
  a cooling flow will develop a cooling flow after the merger.
  This last effect will not be studied in the present work.

The primary role of the radio outburst event (or the merger event) is
to heat the cooling flow, especially in the outer regions, effectively
reducing its age.
 In the present work, we only study
the effects of shocks passing through the cooling flow region.
As we see below, the shock should have velocities above those expected
from the merger process itself.
Supernovae are not likely to supply enough energy to alter a
well developed cooling flow cluster (Wu, Fabian, \& Nulsen 2000),
and we are therefore left with strong radio jets.
{{{ Using the [OIII] luminosities given in Figure 1 of Bicknell {\it et al.}
(1998) along with their equation (1) (or their radio powers and
conversion factors given in Figure 1) we find 
that the kinetic energy (power) of the strongest radio jets are
$\gtrsim 10^{46} \erg \s^{-1}$.
 In total, $\sim 1 \%$ of the systems plotted in Figure 1 of
Bicknell {\it et al.} (1998) have a mechanical power of
$> 10^{46}$ erg s$^{-1}$, and $\sim 10 \%$ have a mechanical jet
power of $> 3 \times 10^{45} \erg \S^{-1}$. }}}
 Operating for $10^7 - 10^8 \yrs$, a jet having a power of
$\sim 10^{46-47}$ erg s$^{-1}$ can heat a total of
$\sim 10^{12} M_\odot$ of gas to a temperature of $\sim 10^{8} \K$.
This is about the mass of gas inside a cooling radius of
a ``standard'' cooling flow with an assumed age of $10^{10} \yrs$ and
a cooling rate of 100 $M_\odot \yr^{-1}$.
 Note that the jet is not required to propagate into the entire cooling
flow region.
{{{ It should be noted that a spherical shock expanding outward
will be slowed down to $\sim 1,000 \km \s^{-1}$ at $\sim 100 \kpc$,
with the energy and density profiles we are using in this section
(this can be shown directly from the self-similar
solution of Chevalier 1976; his eq.~4).
 However, our proposed model assumes that narrow jets expand to the
outer regions, and they maintain their high speeds.
 Regions perpendicular to the jet direction are heated by transverse
shocks and by turbulent mixing.
 These processes require 2D gasdynamical numerical simulations.
In addition, it is quite plausible that the jets precess, hence
expanding along different directions. }}}
 We conclude that very strong radio jets can in principle heat
the gas within $\sim 100 \kpc$ to $\sim 10^8 \K$.

The primary assumption in our scenario is that
a cooling flow cluster undergoes a very powerful radio outburst (on time 
intervals of $2-4 \times 10^9 \yrs$), which results in the
formation of strong radio jets 
with expansion velocities of several$\times 10^3 \km \s^{-1}$. 
The jets may originate from the central cD galaxy, due to a merger
event or accretion of the cooling intracluster gas, or from a galaxy
within the merging subcluster in a merger event.
In the later case, the jets may be excited by the
merger process, or exist prior to the merger. 
The working surface of the jet will propagate through the intracluster gas at velocities of
several$\times 10^3 \km \s^{-1}$, even if the relativistic material 
inside the jet expands at velocities of $>10,000 \km \s^{-1}$.
Shock waves passing through the medium will increase the temperature,
density, and entropy of the gas.

Although the intracluster medium is likely to be multi-phased,
we assume a smooth cooling flow without mass drop out in our calculations.
The same considerations will hold for a multi-phase medium.
In any case, 2-dimensional, or even 3-dimensional numerical calculations
are needed for a detailed analysis of the dynamic evolution of 
a shock within a cooling flow.
Such a study is beyond the scope of the present paper.
The cooling time $t_{\rm cool}$ of a parcel of gas at a distance $r$ from the cluster center is about equal to its inflow time $r/v(r)$,
where $v(r)$ is its inflow velocity.
The numerical simulations of White \& Sarazin (1987) show that the
inflow velocity, temperature, and density can be approximated by
$$
v(r)\simeq v_c(r/r_c)^{-0.9}; \qquad 
T(r)\simeq T_c(r/r_c)^{0.7}; \qquad 
\rho(r)\simeq \rho_c(r/r_c)^{-1.1},
\eqno(3.1)
$$
where $r_c$ is the cooling radius and $v_c \equiv v(r_c)$, etc.
 
Let as assume that a shock wave, which is excited by 
a strong radio jet,  moves through the cooling flow region at a
supersonic velocity $v_g$.
The shocked region has a cylindrical shape of radius $ b \sim 10 \kpc$.
The entropy of the shocked intracluster gas increases as a result of the shock wave.
After being shocked, the gas cools at a higher rate due to the larger
density and temperature, and its entropy decreases.
At the same time, the gas adiabatically re-expands to approximately the same 
pressure
it had before being shocked.
Its final entropy can be greater, equal to, or less than its
value before the shock. We now examine the situation by simple estimates.

 Using the temperature profile in equation (3.1)
the Mach number is 
$$
M_g \equiv {{v_g} \over {c_s(r)}} =  1.5
\left( {{v_g} \over {2,000 \km \s^{-1} }} \right)
\left( {{r} \over {r_c}} \right)^{-0.35},
\eqno(3.2)
$$
where $c_s$ is the adiabatic sound speed and
$T_c=8 \times 10^7 \K$ (White \& Sarazin 1987). In all 
calculations below, we set $\gamma=5/3$ for the adiabatic index of 
intracluster gas.
 From the change of the pressure and density across
a shock we find (e.g., Landau \& Lifshitz 1986, $\S 85$)
$$
{{(P \rho^{-5/3})_2} \over { (P \rho^{-5/3})_{i}}} =
{{5 M_g^2-1} \over {4}}
\left( {{M_g^2+3} \over {4 M_g^2}} \right)^{5/3},
\eqno(3.3)
$$
where a subscript `2' refers to the post shock gas, and `$i$' to the
initial state (the undisturbed cooling flow).
The energy equation in terms of the gas entropy can be written as
$$
{{3} \over {2}} P
{{d} \over {dt}} {\rm ln} P \rho^{-5/3} = -L,
\eqno(3.4)
$$
where $L$ is the cooling rate in units of energy per unit time per unit volume.
We approximate the cooling as $L=\kappa \rho^2  T^{1/2}$.
Therefore, the cooling rate of the post-shock gas is
$L_2=L_i (\rho_2/\rho_i)^2 (T_2/T_i)^{1/2}$. 
In a steady state cooling flow the inflow time is about equal
to the cooling time:  $r/v \simeq (5/2) P_i /L_i$.
Substituting $L_2$ in equation (3.4) with the above expression for
$L_i$, we find 
$$
{{3} \over {2}} P
{{d} \over {dt}} {\rm ln} P \rho^{-5/3} =
-
{{5} \over {2}}
{{v} \over {r}}
\left( {{\rho_2} \over {\rho_i}} \right)^2 
\left( {{T_2} \over {T_i}} \right)^{1/2} 
P_i.
\eqno(3.5)
$$
The effective duration of the post-shock cooling period is the
expansion time of the compressed region, $t_{\rm exp} \simeq b/c_{s2}$,
where $c_{s2} = c_{si} (T_2/T_i)^{1/2}$ is the sound speed of the
post-shocked gas.
 Substituting the effective cooling time in equation (3.5) and integrating,
assuming a constant cooling rate, gives, at the end of the post-shock
radiative cooling period
$$
{{ (P \rho^{-5/3})_{f}} \over { (P \rho^{-5/3})_{2}}} \simeq
\exp \left[
- {{5} \over {3}}
{{v} \over {c_{si}}} {{b} \over {r}}
\left( {{\rho_2} \over {\rho_i}} \right)^{2} 
\left( {{P_2} \over {P_i}} \right)^{-1} 
  \right].
\eqno(3.6)
$$

Combining equation (3.3) with (3.6) the 
ratio of the final to initial values of the function $P/\rho^{5/3}$ is

$$
\eta \equiv {{ (P \rho^{-5/3})_{f}} \over { (P \rho^{-5/3})_{i}}} \simeq
{{5 M_g^2-1} \over {4}}
\left( {{M_g^2+3} \over {4 M_g^2}} \right)^{5/3}
\exp \left[
- {{5} \over {3}}
{{v} \over {c_{si}}} {{b} \over {r}}
\left( {{\rho_2} \over {\rho_i}} \right)^{2} 
\left( {{P_2} \over {P_i}} \right)^{-1} 
  \right].
\eqno(3.7)
$$
The post shock density and pressure are found from the shock conditions
for a Mach number $M_g$, calculated with equation (3.2).
We need to specify the size of the shock wave $b$, and the
shock velocity $v_g$. 
We take the temperature, from which we find $c_{si}$, and the
velocity from equation (3.1) according to White \& Sarazin (1987)
with $r_c = 100 \kpc$, $v_c= 10 \km \s^{-1}$, and $T_c=8 \times 10^7 \K$,
and $b=10 \kpc$
{{{ (the expected size of the cross section of the
shocked intracluster gas around a propagating jet). }}}

If the gas re-expands to its initial pressure, then the final density
and temperature are $\rho_f \simeq \eta^{-3/5} \rho_i$, and
$T_f \simeq \eta^{3/5} T_i$, respectively.
The ratio of the final to initial cooling time is
$$
\left( {{t_f }\over{ t_i }} \right)_{\rm cool}
\simeq \left( {{\rho_f }\over{ \rho_i}} \right)^{-1}
\left( {{T_f }\over{ T_i}} \right)^{1/2} \simeq
\eta^{9/10}
\eqno(3.8)
$$
 The age of the cooling flow is $t_c=r_c/v_c$, and the cooling time at a radius
$r$ is $t_i = t_c (r/r_c) (v/v_c)^{-1}$. Substituting for
$t_i$ in equation (3.8) we find that the ratio of the final cooling time
to the age of the cooling flow is
$$
\beta(r) \equiv \left( {{t_f }\over{ t_c }} \right)_{\rm cooling}
\simeq \eta^{9/10} 
\left( {{r }\over{ r_c }} \right)
\left( {{v }\over{ v_c }} \right)^{-1}. 
\eqno(3.9)
$$
 The ratio $\beta$ is plotted in Figure 1 as a function of the
radius from the cluster center for 3 values of the shock velocity
$v_g$: $3,000$ (solid), $5,000$ (dashed), and $7,000 \km \s^{-1}$
(dash-dot line).

We note the following: (1) The exponential term in equation (3.7) for
$\eta$ has a very small contribution to the results,
since the post-shock cooling time is much longer than
the expansion time under our assumptions.
(2) The cooling time in the outer regions of the cooling flow
can be increased to values much greater
than the age of the cluster, but,
(3) only for high shock velocities $v_g \gtrsim 4,000 \km \s^{-1}$.
Such velocities are appropriate for radio jets, but not for 
merging subclusters. 
Therefore, the cooling flow in the outer region is only terminated
if strong radio jets are formed.
{{{
Carilli, Perley, \& Harris (1994) find that the radio hot spots in the
jets of Cygnus A, a classic Fanaroff-Riley II radio source, have
a Mach number of $\sim5$.
The ambient cluster temperature is such that the associated shock velocity
is then $\sim4,000$ km s$^{-1}$, sufficient to suppress the outer parts
of a cooling flow in this cluster.
}}}

The main conclusion of this section is that a very strong
radio outburst in a cluster can turn-off a cooling flow at radii
beyond $r \gtrsim 0.2-0.5 r_c$.
 This is partially supported by the presence of short radio jets in
some cooling flow clusters.
 We also conclude that the inner regions of cooling flows are
very robust, as the cooling time remains short even after 
the shock propagates through the cluster. 

\section {ACCRETION RATES FROM IMAGING}

Operationally, a cluster is said to have a cooling flow if its central cooling 
time is less than the assumed age $\tau$ of the hydrostatic system of gas.
More specifically, 
the cooling radius $r_c$ (the maximum extent of the cooling flow)
is usually taken to be that radius where the local cooling 
time $t_c$ is equal to the assumed age of the system.
The instantaneous local cooling time is
$$
t_c = \chi{kT\over n\Lambda}, 
\eqno(4.1)
$$
where $n$ is the total particle number density of the gas, $T$ is the gas
temperature, $n^2\Lambda(T)$ is the total 
cooling rate in erg s$^{-1}$ cm$^{-3}$, and $\chi$ is a coefficient of order
unity which reflects uncertainty in the relevant cooling time 
criterion (for isobaric cooling, $\chi=2.5$; for isochoric cooling, 
$\chi=1.5$; there are further differences if the integrated, rather than
instantaneous, cooling time is more appropriate).
Thus, given an observed temperature, the cooling flow condition $t_c=\tau$ 
implicitly determines the cooling radius by defining the density 
$n_c$ at the cooling radius:
$$
n_c\equiv n(r_c)=\chi{kT\over \Lambda\tau} .
\eqno(4.2)
$$
For typical values of $T$ (observed) and $\tau$ (assumed), 
$$
n_c = 0.011
\left( {\chi\over 2.5} \right)
\left( {T\over 4~{\rm keV}} \right)
\left( {\Lambda\over 2\times10^{-23}{\rm ~erg~s^{-1}~cm^3}} \right)^{-1}
\left( {\tau\over 10^{10}{\rm ~yr}} \right)^{-1}  {\rm ~cm^{-3}}. 
\eqno(4.3)
$$

To find the cooling radius $r_c$, the density distribution must be 
derived from the volume X-ray emissivity $\epsilon_X(r)$, which in turn 
must be deprojected from the X-ray surface brightness profile $S_X(r)$.
Outside the cooling flow, the volume emissivity is
$$
\epsilon_X(r)=n^2(r)\Lambda_X(T(r)),
\eqno(4.4)
$$
where $\Lambda_X(T)$ is the amount of cooling in the observed X-ray band.
Inside the cooling flow there is an additional term due to the emission
from cooling flow condensates (White \& Sarazin 1987).
For surface brightness profiles of the form
$$
S_X=S_{X_0} f^{-\sigma}(r/r_0) ,
\eqno(4.5)
$$
where the function $f(r/r_0) = r/r_0$ or $(1 + r^2/r_0^2)^{1/2}$, 
the deprojected emissivity is
$$
\epsilon_X(r) = {1\over \sqrt \pi} 
{\Gamma ( { {\sigma+1} \over 2} ) \over \Gamma( {\sigma\over 2} ) }
{ S_{X_0} \over r_0} f^{-(\sigma+1)},
\eqno(4.6)
$$
where $\Gamma$ denotes the conventional gamma function.
Thus, equations (4.4) and (4.6) imply that the density distribution from 
the cooling radius outward is
$$
n(r)=n_0 f^{-(\sigma+1)/2},
\eqno(4.7)
$$
where
$$
n_0 \equiv \left[ {1\over \sqrt \pi} 
{\Gamma( { {\sigma+1} \over 2} ) \over \Gamma( {\sigma\over 2} ) }
{ S_{X_0} \over r_0 \Lambda_X } \right]^{1/2}.
\eqno(4.8)
$$
The cooling radius is derived implicitly from $n(r)=n_c$, 
where $n(r)$ and $n_c$ are given in equations (4.7-4.8) and (4.2),
respectively.
Adopting a power-law X-ray surface brightness profile for the sake
of illustration, i.e., adopting $f(r/r_0)=r/r_0$ instead of 
$f(r/r_0)=\sqrt{ 1+r^2/r_0^2}$, we see that the cooling radius is given by
$$
r_c = r_0 \left[ \left( {1\over \sqrt \pi}
{\Gamma( { {\sigma+1} \over 2} ) \over \Gamma( {\sigma\over 2} ) }
{S_{X_0} \over r_0 \Lambda_X }\right)^{1/2}
{\tau \Lambda \over \chi k T} \right]^{2/(\sigma+1)} .
\eqno(4.9)
$$
For a power-law X-ray surface brightness profile, $r_0$ is simply some 
fiducial radius where the surface brightness is specified.
Thus, $r_0$ scales with uncertainties in the distance to the cluster,
so $r_0 \propto h^{-1}$.
The cooling radius then scales with the key uncertain variables $h$, 
$\tau$ and $\chi$ as 
$$
r_c \propto h^{-{\sigma \over \sigma +1}} 
\left ( { \tau / \chi }  \right)^{2 \over \sigma + 1} ,
\eqno(4.10)
$$
subject to the constraint $\tau\le H_0^{-1}$.  
If the age is taken be a fraction $\alpha$ of the Hubble time $H_0^{-1}$, then
$$
r_c \propto h^{-{{\sigma +2} \over \sigma +1}} 
\left ( { \alpha / \chi }  \right)^{2 \over \sigma + 1} ,
\eqno(4.11)
$$

To derive the accretion rate at the cooling radius, we use the 
steady state energy equation:
$$
{3\over2}nvk { {\rm d} T \over {\rm d} r} - kTv {{\rm d} n \over {\rm d} r}
= -n^2\Lambda .
\eqno(4.12)
$$
Multiplying both sides by $4\pi r^2$ and defining the accretion
rate to be $\dot M = -4 \pi r^2 \rho v$, where $\rho=\mu m n$ is the total
mass density of the gas and $\mu m$ is the mean mass per particle, 
we find that the accretion rate at the cooling radius is
$$
\dot M_c = { {4 \pi r_c^3 n_c^2 \Lambda} 
{\left( {kT\over \mu m}\right)^{-1} 
\left[{3\over 2}\Delta_r T - \Delta_r n\right]^{-1}}},
\eqno(4.13)
$$
where we use $\Delta_r$ to represent the logarithmic derivative with respect
to $r$:
$$
\Delta_r \equiv { {\rm d~ln~~} \over {\rm d~ln~r} }.
\eqno(4.14)
$$
Substituting for $r_c$ (equation 4.9) and $n_c$ (equation 4.2) we then get
$$
\dot M_c =  {
4 \pi \Lambda r_0^3  \left( {1\over \sqrt \pi}
{\Gamma( { {\sigma+1} \over 2} ) \over \Gamma( {\sigma\over 2} ) }
{ S_{X_0} \over r_0 \Lambda_X } \right)^{3\over(\sigma+1)}
\left( { \chi k T \over \Lambda \tau }\right)^{2(\sigma-2)\over (\sigma+1) }
 { { \left( kT\over \mu m \right)^{-1} } 
\left[{3\over 2} \Delta_r T - \Delta_r n\right]^{-1} } } .
\eqno(4.15)
$$
Thus, the accretion rate at the cooling radius scales as
$$
\dot M_c \propto h^{-{3\sigma\over {\sigma+1}}} 
\left( {\tau / \chi} \right)^{2(2-\sigma)\over(\sigma+1)} .
\eqno(4.16)
$$
Evidently, when $\sigma=2$ (i.e. $S_X\propto r^{-2}$, which is 
typical of cool clusters), 
the inferred accretion rate is independent of the adopted age $\tau$
(provided of course that $\tau \le H_0^{-1}$), but $\dot M_c \propto h^{-2}$.
Again, if $\tau$ is not chosen independently of $H_0$, but taken to
be some fraction $\alpha$ of the Hubble time, then
$$
\dot M_c \propto h^{-{{\sigma+4}\over {\sigma+1}}} 
\left( {\alpha / \chi} \right)^{2(2-\sigma)\over(\sigma+1)} .
\eqno(4.17)
$$

 Despite the fact that equation (4.15) for the evolution of the accretion rate
was derived under the assumption of quasi-hydrostatic equilibrium
near the cooling radius, we believe it provides a reasonable description.
First, for the case $\sigma=1$, the density profile we find from 
equation (4.7) is $\rho \propto r^{-1}$ and the time dependence from 
equation (4.15) is $\dot M_c \propto \tau$. 
  For time dependent self-similar cooling flow solutions, Chevalier (1987) 
assumes isothermal gas and finds a density profile of $\rho \propto r^{-1}$
and obtains $\dot M \propto \tau$,
{{{ as we find here for the same density profile.
 This further strengthen the validity of equation (4.15). 
Second, in the more general numerical study of cooling flow evolution
by Lufkin, Sarazin, \& White (2000), the evolution indicated
by equation (4.15) was found to be qualitatively correct. }}}

  Wu {\it et al.} (2000) find in their calculations of hierarchical
formation of clusters, that the density in the inner regions is
shallower than $\rho \propto r^{-1}$, which means even lower mass
accretion rate in the past.
  Therefore, it is possible that the exact numerical value of the coefficient
in equation (4.15) is not accurate, but the time dependence
may be accurate. 
  If the density profile is not a power law, the behavior is more 
complicated.
  If for example the density profile is
$\rho \propto (1+r^2/R_0^2)^{-(\sigma+1)/2}$, then the dependence on
time will be even stronger at early stages, when $r_c \ll R_0$. 
 We should note that when the density profile changes, so does the
term $\Delta_r n$ in the denominator of equation (4.15).

 The main conclusion is that the accretion rate {\it increases}
with time when the cooling radius is in a region much
smaller than the cluster's core radius (White 1988).  
 We return to the approximation of a power law density profile
[$f=r/r_0$ in equation (4.7)] which was used in deriving equation (4.17).
  In the very inner region (not too close to the central black hole
or other mass concentration) we expect $-1 \leq \sigma < 1$,
so that the cooling increases as $\sim t^4$ or faster.
  If the cooling flow is young, we can safely take, therefore,
$\dot M \propto t^\delta$, so that the total accreted mass is 
$M_{\rm acc} = \dot M_o t_o (t_e/t_o)^{\delta+1}/(\delta+1)$, with 
$\delta \equiv 2 (2-\sigma)/(\sigma+1) >1$,
and where $\dot M_o$ is the cooling rate the cluster would have had 
if the cooling flow age was equal to the current age of the cluster,
$t_o$ is the cluster's age, and $t_e$ is the effective age after
a merger event (or other disruption of an earlier cooling flow). 

 The cooling flow in our model was disrupted a few times
during the cluster evolution, with a typical duration between disruptions of 
$t_e$. 
 We can assume that the number of times it was on is $\sim t_o/t_e$
 We find that the ratio of the total accreted mass in our model to that
of the classical model is 
$$
\left( {M_{\rm acc}}\over{M_{\rm classic}} \right) = 
\left( {1}\over{\delta+1} \right)   
\left( {t_e}\over{t_o} \right) ^\delta  .
\eqno(4.18)
$$
 For reasonable values of $\delta$ and $t_e$, i.e.~$\delta \simeq 2$
(closer to 1 for $\sigma=1$ than to $4$ for $\sigma=0$),
and $t_e=t_o/3$, we get total accreted mass of $1/27$
the classical one. 
The cooling radius will be only $\sim 1/3$ of that for an assumed old
cooling flow, $\sim 50 \kpc$,
assuming an average of $\sigma=1$ in equation (4.9).

\section {DISCUSSION}

  In a recent paper AFJAN, presented a thorough study of seven cooling
flow clusters, using results from ASCA and ROSAT, and argued that these
results support the standard cooling flow model, although with somewhat
younger cooling flow ages.  Although their results strongly support
the presence of cooling flows in seven clusters, we suggest that more moderate
cooling rates are plausible. We consider the following points:
\newline
(1) AFJAN find  that the ages of the seven cooling flows are much 
shorter than the clusters' ages (see also Allen \& Fabian 1997).
The cooling flow ages are in the range $2-7$ Gyr.
 We take this as strong support for our moderate cooling flow model.
 We further argue that relaxing some assumptions
will give even lower ages. 

(2) For the formation of strong radio jets, our model assumes that
most cooling flow clusters harbor a massive black hole,
$M \gtrsim 10^9 M_\odot$, at their centers.
 In M87, the central galaxy of the Virgo cluster,
the mass of the central black hole is estimated to be
$2.5 \times 10^9 M_\odot$ (Harms {\it et al.} 1994),
while the mass of the central black holes in the center
of the cooling flow cluster Abell 2199,
and the galaxies NGC1399 and NGC4472, which are at the center of 
groups with cooling flows, are $\sim 3 \times 10^{10} M_\odot$,
$\sim 5 \times  10^9 M_\odot$, and $\sim 2.6 \times 10^9 M_\odot$, 
respectively (Magorrian {\it et al.} 1998).
 These masses are compatible with our scenario.
AFJAN noted that it is possible that all seven clusters in their
study have central black holes with masses of $\gtrsim 10^{10} M_\odot$.

(3) Even in our moderate cooling flow model, where we find
lower cooling rates, a substantial fraction of the cooling
mass is still unseen.
 However, the fraction of accreted gas that remains unaccounted
for is far higher ($>90\%$) in the standard model.
 Another possible scenario is that of semi-periodic star formation,
in which mass is accumulated until stars form in a weak star-burst
event (McNamara 1999).

(4) The most severe disagreement between AFJAN (and most earlier papers
on the subject) and our suggested model is with regard to the cooling rate.
 AFJAN cite the good match between mass cooling rates deduced from
both X-ray imaging and X-ray spectroscopic analysis to
support the high mass cooling rates.
  We argue that the uncertainties involved in
deducing the mass cooling rates are large, and require strong assumptions.
 Relaxing these assumptions may lead to much lower cooling rates.
 This has been demonstrated by us for the method of X-ray images
in the previous section, where a younger cooling flow gives much lower
mass accretion rate. 
 We turn now to consider mass cooling rates as deduced from
spectral analysis.
\newline
(4.1) The two temperature spectral models used by AFJAN provide a slightly
better description of the ASCA spectra than the cooling-flow models.
This suggests that not all the inferred cooling gas is actually cooling.
As noted by AFJAN (and more references therein), heating by, e.g.
central radio sources, may reduce the inferred cooling rate.
\newline
(4.2) Markevitch {\it et al.} (1998, 1999) using ASCA data,
derive temperature for four clusters in common with AFJAN,
and find temperatures higher by $\sim 15-20$ per cent than those
derived by AFJAN.
Equation (4.15) shows that higher temperatures
lead to smaller accretion rates.
\newline
(4.3) In a joint analysis of {\sl Ginga} LAC and {\sl Einstein} SSS
spectroscopy of clusters, White et al (1994) showed that spectroscopically
deduced accretion rates were uncertain by factors of $\sim2-3$,
{{{ due to uncertainties in the spectrum of cooling flow condensates
and in the amount of intrinsic absorption due to cool gas. }}}
\newline
(4.4)
In deriving the spectral mass cooling rates AFJAN take the mass
deposition rate from the cooling flow to be $\dot M \propto r^q$,
where $r$ is the radial distance from the cluster's center
(their section 3.2).
 They consider $0.75 \leq q \leq 2.75$, where they find the mass cooling
 rate to increase with $q$.
  We examined the dependence of the cooling rate on $q$ as given in their
Table 5, and find that the cooling rate decreases very quickly  with decreasing
values of $q$ below $0.75$.
 According to the proposed model, no mass is cooling to low temperatures
beyond a cooling radius which is much {\it smaller } than the standard
cooling radius.
 This means a very low effective value of $q$ (even negative values !).
The bottom line here is that a much lower mass cooling rate can be
compatible with the spectral observations, $\sim 10 - 30 \%$ of the
accretion rates derived by AFJAN, if lower values of $q$ are allowed.

{{{ Early Chandra observations of Hydra A show that there is a 
significant discrepency between the mass deposition rate
derived from a spectroscopic analysis with that derived from an
image deprojection (David etal 2000). Only the gas within the central 30kpc appears to 
be multiphase with a spectroscopic $\dot M$ a factor of 10 less than 
that inferred from the surface brightness profile.  Hydra A hosts a 
powerful FR Type I radio source and the Chandra observation 
provides strong support for our proposed model.
Over the next few years a large number of cooling flow
clusters will be observed by Chandra, from which we can test 
our proposed moderate cooling flow scenario. }}}

 To summarize this section, we find the disagreement between our proposed
moderate cooling flow model and the spectral mass cooling rate as found
by AFJAN (and earlier papers), to be less embarrassing
than the inability of the standard model to account
for the ultimate sink of cooled mass.
 We argue that relaxed assumptions will lead to much lower cooling rates,
 $\ll 100 M_\odot \yr^{-1}$, both in the 
spectral analysis and imaging analysis, and the much lower total
accreted mass can be accounted for mostly by (possibly cyclical)
star formation and accretion into the central black hole.
 Hence, only a mass of $\ll 10^{10} M_\odot$ presently resides
in a cool phase ($T \lesssim 10^4 \K$).

\section {SUMMARY}

 The goal of the present paper is to make several major
observational properties of cooling flow clusters more consistent:
most clusters contain cooling flows, which contain signs of
strong activity in their very inner regions, but
the lack of an observed reservoir for the large mass of gas
that has cooled over a Hubble time is especially problematic for the
``standard model''.
 We propose that the effective ages of cooling flows are much
lower than cluster ages.
 The effective age of a cooling flow is the time since it was last
largely disrupted by 
strong radio jets, which were triggered by a subcluster merger 
or by the accretion of cooling intracluster gas.
 The main assumption of the proposed scenario is that, on a time scale of
$\sim 2-4 \times 10^9 \yrs$ the central black hole of a cooling flow
cluster generates a very powerful radio outburst, resulting in strong
radio jets that expand at velocities of several$\times 10^3 \km \s^{-1}$
in the inner regions ($r \lesssim 100 \kpc$) of the cluster. 
 In a merger event, the jets may come from the nucleus of the cD galaxy,
or from one of the galaxies of the merging subcluster that crosses
the central region.
 The jets remain strong for $\sim 10^7 - 10^8$ years.

 The main results of the proposed scenario can be summarized as follows:
(1) The total mass that cools to temperatures below $10^4 \K$ is more 
than an order of magnitude below the mass expected from the standard model
of cluster cooling flows (which assumes an age of $\sim 10^{10} \yrs$ and
a constant mass cooling rate).
(2) Low star formation rates and accretion onto the central black hole
can account for most of the mass that cooled over the life of the
cooling flow.
 Only $\ll 10^{10} M_\odot$ presently resides
in a cool phase ($T \lesssim 10^4 \K$).
(3) The very inner regions of cooling flows $r \lesssim 10-30 \kpc$, are very
robust to destruction, and therefor are long lived.
This explains why all the unique properties of cooling flows
(e.g., optical filaments, blue excess) are found in the inner regions.
(4) This also explains the fact that cooling flows are observed in most
clusters.
(5) The robustness of the inner regions of cooling flow clusters suggests
  that clusters currently without a cooling flow never had one.
(6) The heating of the outer regions by shocks and the robustness of the
inner regions lead us to predict that high spatial resolution observations 
(e.g., with {\it Chandra}) will show a shallower temperature
profile in the outer regions ($r \gtrsim 50 \kpc$), with a steeper fall
of the temperature (moving inward) at $\sim 30-50 \kpc$.
{{{ In addition, due to the jets which once were expanding through these
regions, we predict that the cooling flow regions will show some
strong inhomogeneities, even in regions were there
are presently no radio jets. }}}
(7) The radio activity in cooling flow clusters was much larger in the
past, at a redshift of $z \sim 0.5$.
 Due to the larger density of intracluster gas, the radio jets did not expand
though to large distances from the center. 

{\bf ACKNOWLEDGMENTS:} 
 We thank an anonymous referee for helpful comments.
 We thank Paul Nulsen for helpful discussions.
This research was supported by grant NAS8-39073.
N.S. thanks the CfA for hospitality during three visits
and acknowledges support from an Israel Science Foundation grant and an
Israel-USA Binational Science Foundation grant.
R.E.W.  acknowledges partial support from NASA grant
NAG 5-2574 and a National Research Council Senior Research Associateship
at NASA GSFC.


{\bf FIGURE CAPTIONS}

\noindent {\bf Figure 1:}

 The ratio of the gas final cooling time to the age of the original
cooling flow as a function of radial distance from the cluster
center, after a passage of a shock wave through the cooling
flow region.
 The ratio $\beta$ (see eq. 3.9) is plotted for 3 values of the shock
velocity $v_g$: $3,000$ (solid), $5,000$ (dashed), and $7,000 \km \s^{-1}$
(dash-dot line).
 The undisturbed cooling flow model is according to equation (3.1)
with a cooling radius of $r_c=100 \kpc$.

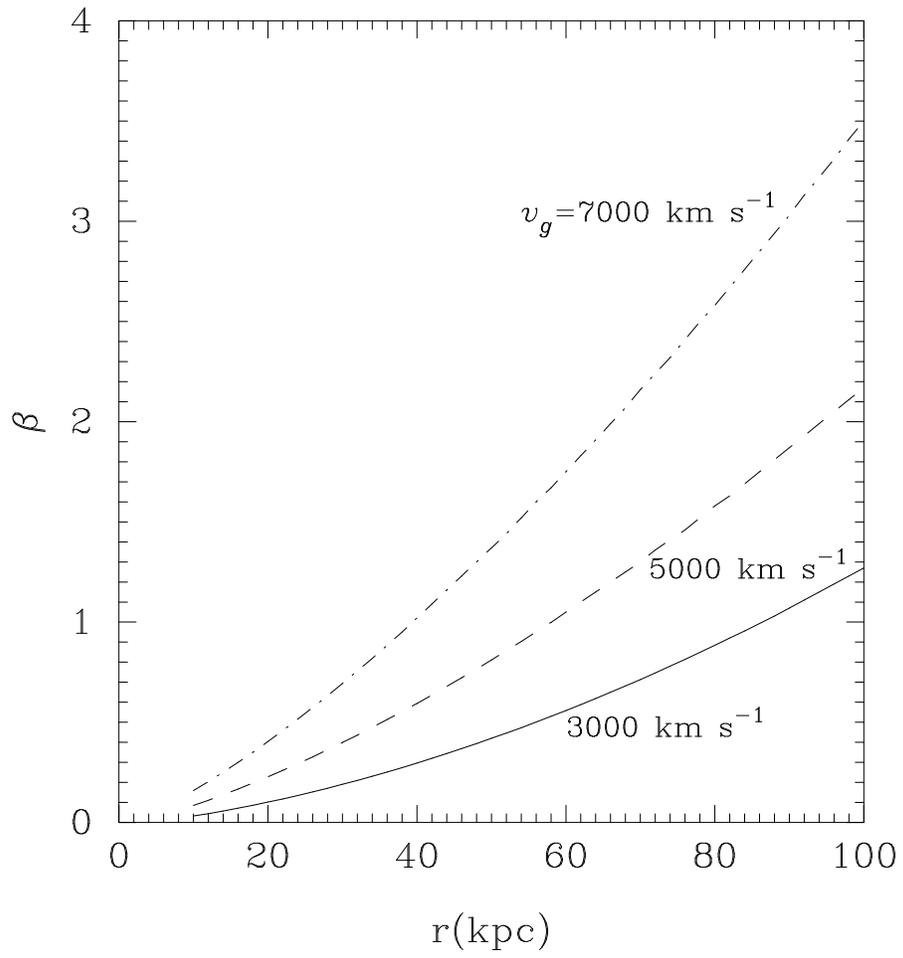
\end{document}